\documentstyle[psfig]{mn}
\newif\ifAMStwofonts
\AMStwofontstrue

\newcommand{\f}{\frac}

\newcommand{\bb}{\bibitem}
\newcommand{\BF}{\begin{figure}\begin{center}}
\newcommand{\EF}{\end{center}\end{figure}}
\newcommand{\BE}{\begin{equation}}
\newcommand{\EE}{\end{equation}}

\ifoldfss
  \ifCUPmtlplainloaded \else
    \NewTextAlphabet{textbfit} {cmbxti10} {}
    \NewTextAlphabet{textbfss} {cmssbx10} {}
    \NewMathAlphabet{mathbfit} {cmbxti10} {} 
    \NewMathAlphabet{mathbfss} {cmssbx10} {} 
  \fi
  \ifAMStwofonts
    \ifCUPmtlplainloaded \else
      \NewSymbolFont{upmath} {eurm10}
      \NewSymbolFont{AMSa} {msam10}
      \NewMathSymbol{\upi}     {0}{upmath}{19}
      \NewMathSymbol{\umu}     {0}{upmath}{16}
      \NewMathSymbol{\upartial}{0}{upmath}{40}
      \NewMathSymbol{\leqslant}{3}{AMSa}{36}
      \NewMathSymbol{\geqslant}{3}{AMSa}{3E}

      \let\leq=\leqslant \let\le=\leqslant
       \let\ge=\geqslant
    \fi
  \fi
\fi 

\ifnfssone
  \newmathalphabet{\mathit}
  \addtoversion{normal}{\mathit}{cmr}{m}{it}
  \addtoversion{bold}{\mathit}{cmr}{bx}{it}
  \newmathalphabet{\mathbfit} 
  \addtoversion{normal}{\mathbfit}{cmr}{bx}{it}
  \addtoversion{bold}{\mathbfit}{cmr}{bx}{it}
  \newmathalphabet{\mathbfss} 
  \addtoversion{normal}{\mathbfss}{cmss}{bx}{n}
  \addtoversion{bold}{\mathbfss}{cmss}{bx}{n}
  \ifAMStwofonts
    \ifCUPmtlplainloaded \else
      \UseAMStwoboldmath
      \makeatletter
      \new@mathgroup\upmath@group
      \define@mathgroup\mv@normal\upmath@group{eur}{m}{n}
      \define@mathgroup\mv@bold\upmath@group{eur}{b}{n}
      \edef\UPM{\hexnumber\upmath@group}
      \new@mathgroup\amsa@group
      \define@mathgroup\mv@normal\amsa@group{msa}{m}{n}
      \define@mathgroup\mv@bold\amsa@group{msa}{m}{n}
      \edef\AMSa{\hexnumber\amsa@group}
      \makeatother
      \mathchardef\upi="0\UPM19
      \mathchardef\umu="0\UPM16
      \mathchardef\upartial="0\UPM40
      \mathchardef\leqslant="3\AMSa36
      \mathchardef\geqslant="3\AMSa3E

      \let\leq=\leqslant \let\le=\leqslant
       \let\ge=\geqslant
    \fi
  \fi
\fi 

\ifnfsstwo
  \DeclareMathAlphabet{\mathbfit}{OT1}{cmr}{bx}{it}
  \SetMathAlphabet\mathbfit{bold}{OT1}{cmr}{bx}{it}
  \DeclareMathAlphabet{\mathbfss}{OT1}{cmss}{bx}{n}
  \SetMathAlphabet\mathbfss{bold}{OT1}{cmss}{bx}{n}
  \ifAMStwofonts
    \ifCUPmtlplainloaded \else
      \DeclareSymbolFont{UPM}{U}{eur}{m}{n}
      \SetSymbolFont{UPM}{bold}{U}{eur}{b}{n}
      \DeclareSymbolFont{AMSa}{U}{msa}{m}{n}
      \DeclareMathSymbol{\upi}{0}{UPM}{"19}
      \DeclareMathSymbol{\umu}{0}{UPM}{"16}
      \DeclareMathSymbol{\upartial}{0}{UPM}{"40}
      \DeclareMathSymbol{\leqslant}{3}{AMSa}{"36}
      \DeclareMathSymbol{\geqslant}{3}{AMSa}{"3E}

      \let\leq=\leqslant \let\le=\leqslant
       \let\ge=\geqslant
    \fi
  \fi
\fi 

\ifCUPmtlplainloaded \else
  \ifAMStwofonts \else 
    \def\upi{\pi}
    \def\umu{\mu}
    \def\upartial{\partial}
  \fi
\fi

\begin{document}
\title{Temperature Correlations in a Compact Hyperbolic Universe}
\author[Kaiki Taro Inoue, Kenji Tomita and Naoshi Sugiyama]
{Kaiki Taro Inoue$^{1}$, Kenji Tomita$^{1}$ and Naoshi Sugiyama$^{2}$
\\
$^{1}$Yukawa Institute for Theoretical Physics, Kyoto University, 
Kyoto 606-8502, Japan.\\ $^{2}$Department of Physics, 
Kyoto University, Kyoto 606-8502, Japan}

\maketitle

\date{15 August}

\begin{abstract}
The effect of a non-trivial topology on the temperature 
correlations on the cosmic microwave background (CMB) in a small compact
hyperbolic universe with volume comparable to the cube of the
curvature radius is investigated.  
Because the bulk of large-angular CMB fluctuations is produced at
the late epoch in low $\Omega_0$ models, the
effect of a long wavelength cut-off due to the periodic structure
does not lead to the significant suppression of large-angular power 
as in compact flat models.  The angular power spectra are consistent with the COBE data for $\Omega_0\!\ge\!0.1$. 
\end{abstract}


\section{Introduction}
\indent
Einstein's equations do not specify the global structure
of spacetime. In other words, to a given local metric, a large number of
topologically distinct models remain unspecified. 
In the absence of the unified theory that describes the global
structure as well as the local one, one must resort to the
observational  methods to determine the global topology of the universe.
\\
\indent
Assuming that the spatial hypersurface is homogeneous, 
the observed high degree of isotropy in the cosmic microwave background 
(CMB) points to the Friedmann-Robertson-Walker (FRW) models as the best  
candidates of the cosmological models. 
However, if one would allow the spatial hypersurface being
multiply-connected, a variety of locally FRW models which are globally 
anisotropic and inhomogeneous may be consistent with the 
current observational data.
\\
\indent
Constraints on the topological identification scales using the COBE
data have been obtained for
some flat models with no cosmological
constant (Stevens, Scott \& Silk 1993; de Oliveira, Smoot \& 
Starobinsky 1996; Levin, Scannapieco \& Silk 1998)
 and some limited compact hyperbolic (CH) models (Levin, Barrow, Bunn \& 
Silk 1997; Bond, Pogosyan \& Souradeep 1998).
The large-angular temperature 
fluctuations discovered by the COBE constrain the  
possible number of the copies of the fundamental domain inside 
the last scattering surface to less than $\!\sim$8 for 
compact flat multiply-connected models.
\\
\indent
On the other hand, a large amount of CMB anisotropies on large 
scales could be produced in the low density universe due to the decay of 
gravitational potential near the present epoch (Cornish,
 Spergel \& Starkman 1998).
Therefore we expect that the 
constraint on the possible number of copies 
is less stringent for CH models.
However, since the effect of the 
non-trivial topology
becomes more and more significant as the volume of the space decreases, it is 
very important to investigate the viability of the CH models with 
small comoving volume. 
\\
\indent
From a theoretical point of view, the ''smallness'' of the spatial
hypersurface is an advantage for giving a natural mechanism
leading to homogeneity and isotropy. It is well known that
geodesic flows on CH spaces are strongly chaotic. Therefore, 
initial perturbations would be smoothed out due to the mixing effects
(Lockhart, Misra \& Prigogine 1982; Gurzadyan \&
Kocharyan 1992; Ellis \& Tavakol 1994). In inflationary
scenarios, a certain physical process is indispensable that homogenises the
initial patch beyond the horizon scale before the onset of inflation 
for accomplishing the sufficient smoothing of the observable universe
(Goldwirth \& Piran 1989; Goldwirth 1991). The chaotic mixing in CH
spaces may provide a solution to the pre-inflationary initial value
problem (Cornish, Spergel \& Starkman 1996).
\\
\indent 
If we live in a \textit{small universe} which is defined to be a
locally homogeneous and isotropic space that is multiply-connected
on scales comparable to or smaller than the horizon, the future 
astronomical satellite missions such as MAP and PLANCK might reveal some 
specific features in CMB (Cornish, Spergel, \& Starkman 1998; Weeks 1998).
\\
\indent
So far, a variety of CH manifolds have been constructed by
mathematicians. 
However, the number of the known CH manifolds with small volume is
relatively small. 
In this paper, we investigate CH models whose spatial
hypersurface is isometric to the Thurston manifold which is 
the second smallest in the known CH manifolds with  
volume 0.98139 times cube of the curvature
radius. The smallest
one is the Weeks manifold with volume 96 percent of that of the
Thurston manifold(see e.g. Fomenko \& Kunii 1997). However, the 
fundamental domain(which tesselates the
infinite space) of the Thurston manifold is much simpler than that
of the Weeks manifold. For simplicity, we investigate the 
Thurston models rather than the Weeks models. The fundamental 
domain of the Thurston manifold is a polygon with 16
faces, which can be constructed by appropriately 
identifying 8 faces 
with the remaining 8 faces(see the appendix of Inoue 1999a). 
It should be noted that the volume of CH manifolds
must be larger than 0.16668 times cube of the curvature
radius although no concrete examples of manifolds with such small volumes
are known (Gabai, Meyerhoff \& Thurston 1996).  
\section{Computation of eigenmodes}
\indent
So far various kinds of numerical techniques have been proposed to 
overcome the difficulty of computing the CMB in CH models. 
For several CH models, CMB fluctuations
have been computed using the method of images without carrying out
the mode expansion (Bond, Pogosyan, \& Souradeep 1998).
 They obtained the result that 
the COBE data strongly constrains the CH models so that 
the comoving volume of the 
fundamental domain are at least comparable to the comoving volume inside the
last scattering surface.  Since the method of images requires the sum of 
exponentially increasing images, it is difficult to obtain the distinct
eigenmodes which are necessary to estimate the effect of the power
spectrum with discrete peaks. Alternatively,
one of the author proposed a numerical approach called 
the direct boundary element method for computing eigenmodes of the
Laplace-Beltrami operator (Inoue 1999a). 14 eigenmodes have been computed for 
the Thurston manifold.
It is numerically found that the expansion coefficients behave as 
if they are random Gaussian numbers. 
\\
\indent
In this work, we have numerically computed 36 
eigenmodes in the Thurston manifold up to $k\!=\!13$ 
(the curvature radius is normalized to one) 
which are approximated by quadrature shape 
functions which converges to the solutions faster than
constant valued shape functions.  As we shall see, the 
contribution of the higher modes to the angular power spectra on large
angular scales are relatively small for low-density models. 
In other words, the effect of the non-trivial topolgy is 
almost determined by the lower modes. 
We confirm the previous computed
eigenvalues within $|\delta k|\leq 0.01$.
\\
\indent
We see from figure 1 that the 
number of eigenmodes below $k$ is nicely
fitted to the Weyl's asymptotic formula 
\begin{equation}
N(k)=\f{\textrm{Vol} ({M})(k^2-1)^{3/2}}{6 \pi^2},
~~~k\!>\!>\!1,
\end{equation}
where Vol$(M)$ denotes the volume of a manifold $M$.
The random Gaussian behavior is again observed for 31 modes
$5.404\!\leq\!k\!<\!13$ but five degenerated states have an 
eigenmode which shows the non-Gaussian behavior due to the global
symmetry of the fundamental domain. It is found that the five eigenmodes
have $Z2$ symmetry 
(invariant with respect to the rotation by an angle $\pi$) 
on the center (where the minimum length of the periodic geodesic which
lies on the point is locally maximal) of the fundamental domain. 
In this case, one would
observe an axis around which the fluctuation is rotationally symmetric 
at the center. 
Therefore, the correlation between expansion coefficients
leads to a non-Gaussian behavior. 
Nevertheless, it is found that 
appropriate choices of the linear combination of the 
degenerated modes recover the generic Gaussian behavior.
Furthermore, the symmetry of CH manifolds depends on the 
observing point. If one randomly choose a point on the manifold,
the probability of observing an exact symmetry of the manifold is 
very small. 
The result supports
the previous investigations of the expansion
coefficients which show
the Gaussian behavior in classically chaotic systems
(Aurich \& Steiner 1989; Haake \& Zyczkowski 1990)
 although the global symmetry in the system can hide 
the generic property (Balazs \& Voros 1986).
\section{Temperature Fluctuations}
\indent
Perturbations in CH models can be written in terms of 
linear combination of
eigenmodes on the universal covering space multiplied by the expansion
coefficients and the initial fluctuations plus time evolution of the 
perturbations. The expansion coefficients
include the information of
the periodicity in the universal covering space.
As CH models are locally homogeneous and isotropic, the time evolution
of the perturbations coincides with that in open models.
\BF
\centerline{\psfig{figure=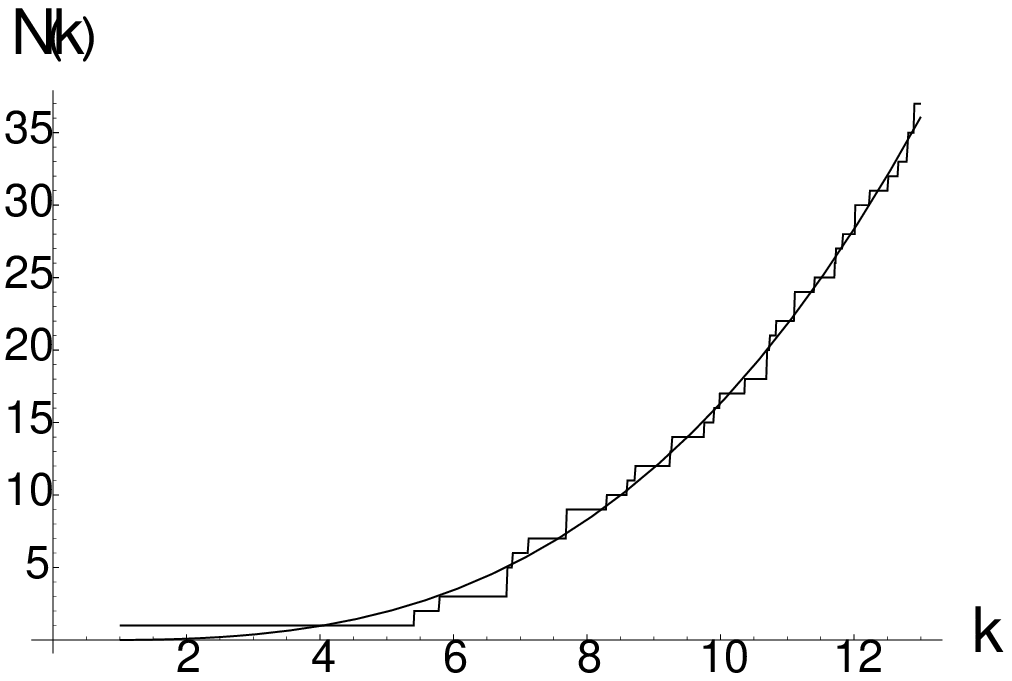,width=7.2cm}}
\caption{Number function and the Weyl's
asymptotic formula.}
\label{fig:Num13}
\EF
\\
\indent
The dominant physical effects producing CMB anisotropies (Hu, Sugiyama \&
Silk 1997)
on large angular scales are the 
ordinary Sachs-Wolfe (OSW) effect (Sachs \& Wolfe 1967), which is 
the gravitational redshift effect in between the last scattering surface 
and the present epoch, 
and the integrated Sachs-Wolfe (ISW) effect, which is 
the gravitational blue-shift effect caused by the decay of 
gravitational potential at the curvature domination epoch, 
$1\!+\! z \!\sim\! (1-\Omega_0) / \Omega_0$.
For the COBE scales, we can ignore the contribution 
from the acoustic oscillations.
Then the time evolution of the 
adiabatic growing mode of 
the Newtonian gravitational potential is analytically given
as (see e.g. Kodama \& Sasaki 1986; Mukhanov, Feldman \& Brandenberger 
1992)
\begin{equation}
\Phi_t(\eta)=\Phi_t(0)
\f{5(\sinh^2 \eta-3 \eta\sinh\eta+4 \cosh\eta-4)}
{(\cosh\eta-1)^3},
\end {equation}  
where $\eta$ denotes the conformal time. 
The two-point
temperature correlations in a CH cosmological model can be
written in terms of the gravitational potential. 
Assuming that the initial fluctuations obey the Gaussian statistic, 
and neglecting the tensor-type perturbations, 
the angular power spectrum $C_l$ can be written as
\begin{eqnarray}
(2\,l+1)\,C_l
&=&\sum_{m=-l}^{l} \langle~ |a_{lm}|^2 \rangle
\nonumber
\\
&=&\sum_{\nu,m}\f{4 \pi^4~{\cal P}_\Phi(\nu) }
{\nu(\nu^2+1)\textrm{Vol}(M)}~|\xi_{\nu l m}|^2 |F_{\nu l}|^2 ,
\end{eqnarray}
where
\BE
F_{\nu l}(\eta_o)
\!\!\equiv\!\!\f{1}{3}
\Phi_t(\eta_\ast) X_{\nu l}(\eta_o\!-\!\eta_\ast)
\!+\!\! 2 \!\!\int_{\eta_\ast}^
{\eta_o}\!\!\!\!\!\!d \eta\, 
\f{d\Phi_t}{d \eta}X_{\nu l}(\eta_o\!-\!\eta).\label{eq:cor}
\EE
Here, $\nu\!=\!\sqrt{k^2-1}$, ${\cal P}_\Phi(\nu) $ 
is the initial power spectrum, and
$\eta_\ast$ and $\eta_o$ are the conformal time of the
last scattering and the present conformal time, respectively. 
$X_{\nu l}$ denotes the radial eigenfunctions in open models and
$\xi_{\nu l m}$ denotes the expansion coefficients. 
 From now on we assume that the initial power spectrum is the
(extended) Harrison-Zeldovich spectrum $i.e.$, ${\cal P}_\Phi(\nu)
\!=\!Const.$.
\\
\indent
Although the low-lying modes give an appreciable contribution to the 
large angular power, contributions of higher eigenmodes may
not completely be negligible.
While the computation of highly-excited eigenmodes is a difficult 
task,  we have so far succeeded to calculate the exact eigenmodes up to 
$k\!=\!13$ as we mentioned before. However,  
we are going to assume that $\xi_{\nu l m}$ 's are 
also random Gaussian numbers for higher modes.
Since the
information of the periodicity in the real space is lost by this
approximation, we will only employ this approximation to 
the statistics in the $k$-space which is expected 
to be not changed because the periodicity 
is not apparent in the $k$-space. 
As CH models are globally
inhomogeneous, the expected correlation statistics depend on the
point of the observer.  Therefore, one can interpret 
that one realization for the expansion coefficients
corresponds to a certain point of the observer in
the fundamental domain.  In order to apply the random Gaussian
approximation, one must also estimate the variance of the expansion
coefficients. The expansion coefficients are written in terms of
eigenmodes $u_\nu$ and spherical harmonics $Y_{lm}$ as
\begin{equation}
\xi_{\nu l m} X_{\nu l}(\chi_o)=\int u_\nu(\chi_o,\theta,\phi)\, 
Y^*_{l m}(\theta,\phi) d \Omega. \label{eq:Intxi}
\end{equation}   
It should be noted that (\ref{eq:Intxi}) is satisfied at arbitrary
radius $\chi_o$. Let us consider a sphere with large radius 
$\chi_o\!>\!>\!1$ on the Poincar$\acute{\textrm{e}}$ ball which is the 
image of the upper hyperboloid in the four-dimensional Minkowski space
($y_0,y_1,y_2,y_3$) by a stereographic projection onto the unit ball 
on the ($0,y_1,y_2,y_3$) plane using a point ($-1,0,0,0$) as the base point. 
One can expect the random behavior of the mode functions on the sphere
as the surface of the sphere which is pulled back by the discrete 
isometry group fills the
fundamental domain ergodically. The (apparent) angular fluctuation
scale $\delta \theta$ of
$k$-mode is approximated in terms of two parameters $\chi_o$ and $k$
as ,
\begin{equation}
\delta \theta^2 \sim \f{16 \pi^2~\textrm{Vol}(M)}  
{k^2 (\sinh(2(\chi_o+r_{ave}))
-\sinh(2(\chi_o-r_{ave}))-
4 r_{ave})}, \label{eq:FS} 
\end{equation}
where $r_{ave}$ denotes the averaged radius of the inradius and
outradius of the fundamental domain.
One can approximate $u_{\nu'}(\chi_o)\sim u_\nu(\chi_o')$ by choosing an
appropriate radius $\chi_o'$ which satisfies
$k^{-2}\exp(-2\chi_o')=k'^{-2}\exp(-2\chi_o)$. 
Averaging (\ref{eq:Intxi}) over $l$ and
$m$, one obtains
\begin{equation}
\bigl <|\xi_{\nu' l m} |^2 \bigr >
\sim 
\f{\exp(-2\chi_o')}{\exp(-2\chi_o)}
\bigl<|\xi_{\nu l m} |^2 \bigr >,
\end{equation}  
which gives $\bigl<|\xi_{\nu l m}|^2 \bigr> \sim \nu^{-2}$. 
We have found that the computed variances of $\xi_{\nu l m}$'s for
$2\!\leq l\!\leq 20$, $-l\!\leq m \!\leq l$ are 
remarkably in good agreement with the analytical estimate. 
\BF
\centerline{\psfig{figure=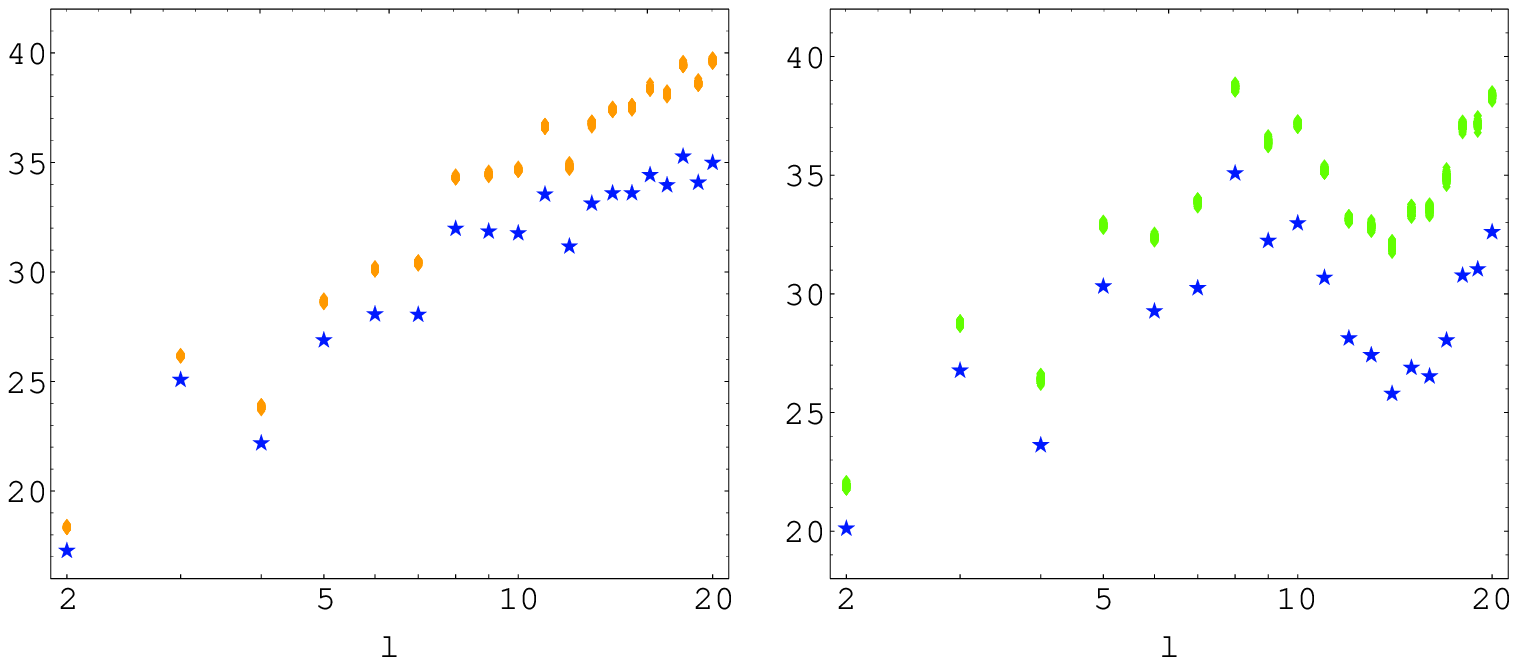,width=9.8cm}}
\caption{$\delta T_{l}/T \equiv 
\sqrt{l(l+1)C_{l}/(2 \pi)}$ for the Thurston models 
$\Omega_{0}\!=\!0.2$ (left) and $\Omega_{0}\!=\!0.4$ (right)  
using expansion coefficients derived from 36 eigenmodes 
only(stars) and that using these coefficients and 
random Gaussian numbers for $13\!<\!k\!<\!50$ 
with 100 realizations (diamonds). Eigenvalues for higher modes
are approximated by Weyl's asymptotic formula.}
\label{fig:RGapp24}
\EF
\\
\indent
From figure 2, one can see that the uncertainty in the Gaussian
approximation is very small.  Remarkably, each realization gives almost the
same value so that 100 points for given $l$ are plotted as a tiny speck.  
The contribution of higher modes becomes significant as $\Omega_0$ is
increased because the curvature dominant era is shifted to the late
time so that the OSW effect becomes 
dominant over the ISW effect.
It is found that contributions of the modes $k\!>\!13$ to $C_{l}$ for 
$2\!\le\!l\!\le\!20$
are approximately $7$ percent and $10$ percent for $\Omega_{0}\!=\!0.2$
and $\Omega_{0}\!=\!0.4$, respectively. Thus contribution of 
modes $k\!>\!13$ which we employ Gaussian approximation
is almost negligible on 
large angular scales especially in low $\Omega_0$ models. 
\BF
\centerline{\psfig{figure=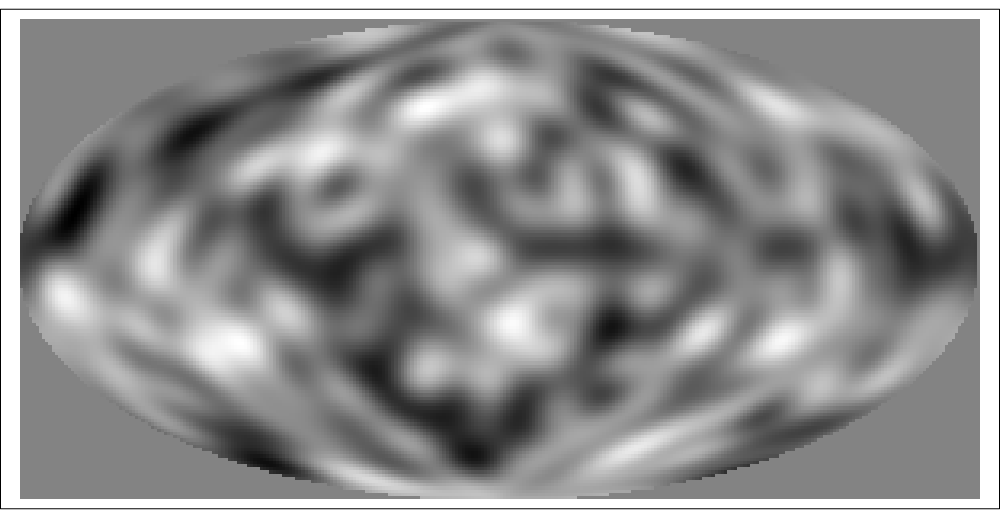,width=8cm}}
\caption{A simulated sky map of the microwave background (convolved 
with the COBE DMR beam) in the Thurston model 
$\Omega_0=0.2$. }
\label{fig:SKY0.2}
\EF
One realization (for the initial fluctuation) 
of a typical CMB fluctuation as seen by COBE is plotted in
figure 3 for $\Omega_0\!=\!0.2$.  In the simulation, 
we used only ''exact'' 36 eigenmodes. We have chosen a point  
where the injective radius is maximal as the center
(belonging to the ''thick'' part of the manifold). 
One can see that the structure 
due to the periodical boundary conditions is not apparent. 
However, approximated number of copies of the
fundamental domain inside the last scattering surface is $\sim 500$
for the Thurston model with $\Omega_0\!=\!0.2$. Therefore, the effect
of the non-trivial topology is expected to be significant. 
\BF
\centerline{\psfig{figure=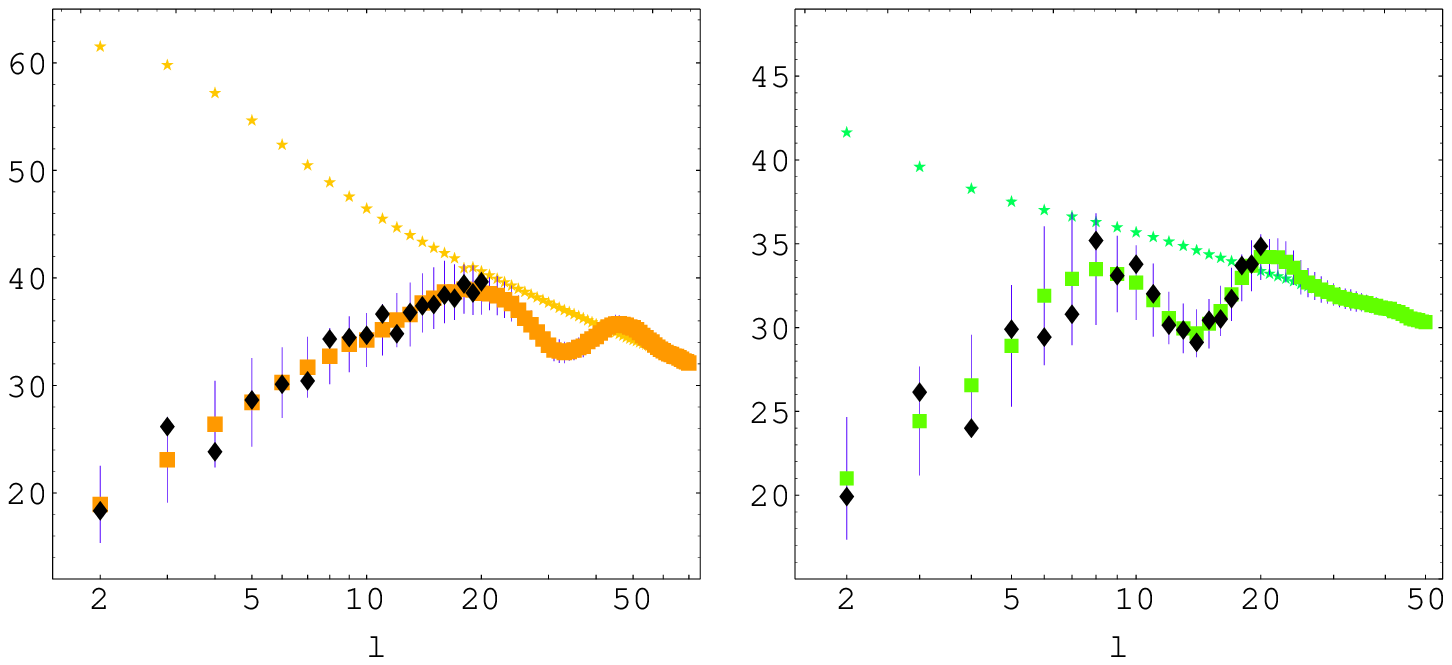,width=9.6cm}}
\caption{$\delta T_{l}/T\!=\!\sqrt{l(l+1)C_{l}/(2 \pi)}$
  for the Thurston models at center (diamonds)
and ensemble averaged values (boxes) and open
models (stars) with $\Omega_0=0.2$ (left) and $\Omega_0=0.4$ (right). Because 
of the global inhomogeneity, $\delta T_{l}/T$'s have dependence on the
observing points inside the fundamental domain that causes the
uncertainty in $\delta T_{l}/T$. Two-sigma ''geometric variance'' is
shown in vertical lines which has been
obtained by $500$ realizations for the expansion coefficients.}
\label{fig:OTL24}
\EF
\\
\indent
The mode cut-off at $k\!=\!5.404$ 
which corresponds to the largest wavelength
inside the fundamental domain
causes the suppression of the
angular power on large angular scales as in compact flat models. 
However, the decay of the
Newtonian potential 
in the curvature dominant era makes the difference.
Since the bulk of the large angular power comes from 
the decay of the potential
well after the last scattering time,  
the large angular power does not suffer 
the significant suppression. We see from figure 4 
that the slope of the large angular power is not
steep even for the model with $\Omega_0\!=\!0.2$ in contrast to the 
compact flat models without cosmological constant. 
The two peaks in the power spectrum for the CH
model are important in understanding the effect of the 
non-trivial topology. 
The angular scale which gives  
the first peak is equivalent to the
angular fluctuation scale of the lowest eigenmode ($k\!=\!5.404$)
on the last scattering surface. 
Substituting the comoving radius of the last 
scattering surface in unit of the curvature radius 
$R_{curv}$,
\BE
R_{LSS}=R_{curv} \cosh^{-1}(2/\Omega_0-1)
\EE
into (\ref{eq:FS})
gives the angular scales $l\!=\!17$ for $\Omega_0\!=\!0.2$ 
and $l\!=\!7.4$ for $\Omega_0\!=\!0.4$. 
Beyond this scale, the OSW contribution is strongly suppressed as 
in compact flat models. However, eigenmodes with angular 
scales below the given scale
at the last scattering can have large angular scales after the last
scattering. Therefore, in the presence of the ISW
effect, the suppression of the power beyond the scale which corresponds 
to the first peak is very weak in contrast to flat models. 
The angular scale which gives the second peak corresponds to 
the scale of the projected lowest eigenmode at the last scattering. 
Below this scale, the angular power asymptotically 
converges to that of open models because the effect of the modes with 
wavelength larger than the cut-off wavelength is negligible. 
Since we have ignored the effects of subhorizon perturbations
at  the last scattering
such as the so-called 'early' ISW effect during the matter-radiation 
equality epoch
and the Doppler effect due to the 
acoustic velocity, the angular 
power on large to intermediate scales must be slightly
boosted. However, these effects are irrelevant to the global effect 
of the non-trivial topology inasmuch as one considers the typical 
topological identification scale that is not significantly smaller than
the present horizon.  
\BF
\centerline{\psfig{figure=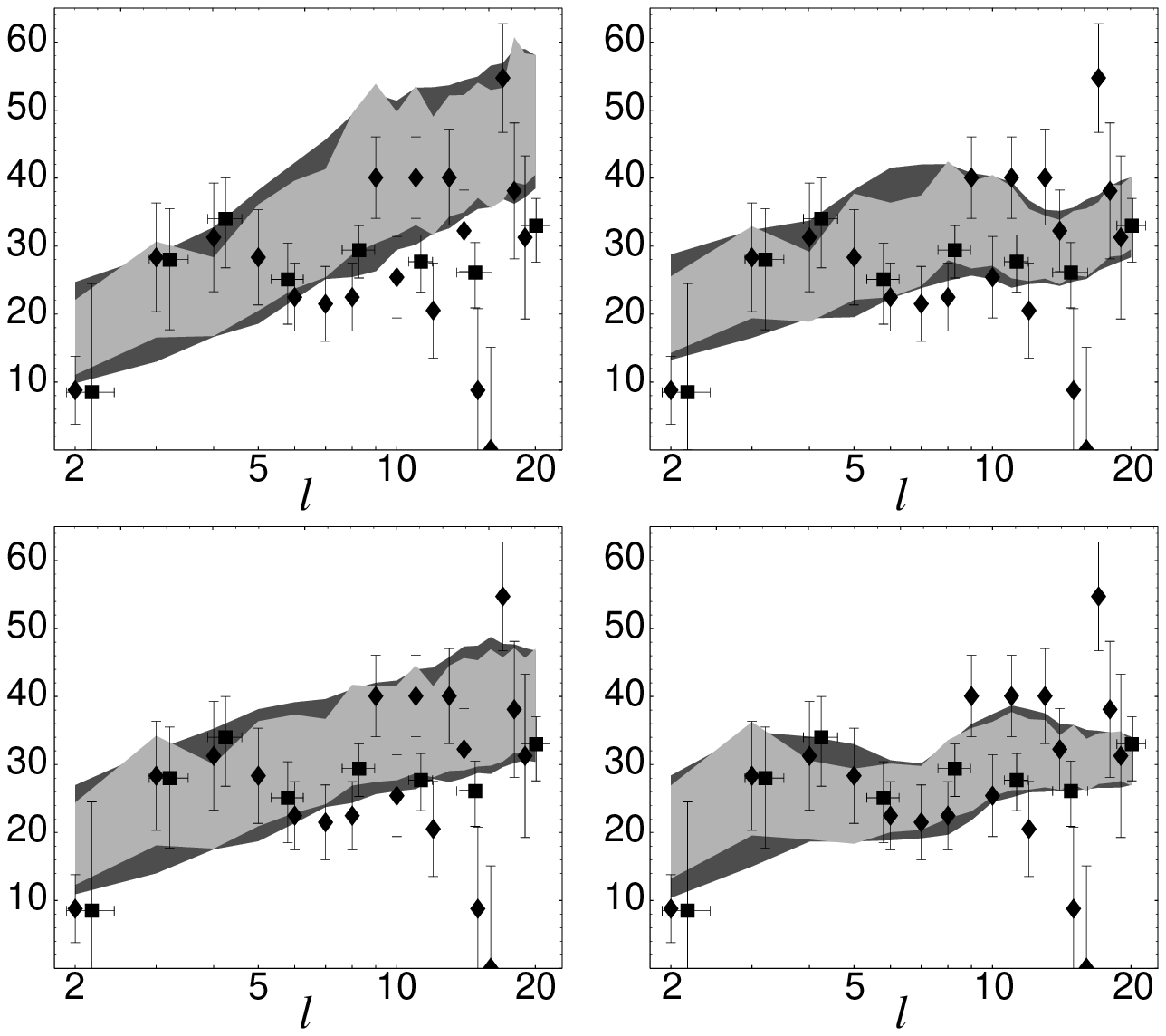,width=9.6cm}}
\caption{$\delta T_{l}/T\!=\!\sqrt{l(l+1)C_{l}/(2 \pi)}$ 
in $\mu\!K$ for the Thurston model with
$\Omega_{0}\!=\!0.1$ (top left) and $\Omega_{0}\!=\!0.4$ (top right)
and $\Omega_{0}\!=\!0.2$ (bottom left) and $\Omega_{0}\!=\!0.6$ (bottom right)
The light-gray band corresponds
to the one-sigma cosmic variance in center and 
the darkgray band corresponds to 
the net variance of the one-sigma cosmic and the 
one-sigma ''geometric'' variances. The COBE DMR measurements analysed 
by Gorski and Tegmark are 
plotted in diamonds and boxes respectively.}
\label{fig:COSMIC1426}
\EF
\\
\indent
In figure 5, the angular power spectra for low-Omega
models are plotted with the COBE data (Gorski et al, 1996) (diamonds).
They have been calculated 
using 36 eigenmodes and the Gaussian approximation 
taking account of $\sim$10 percent contributions 
from higher eigenmodes. 
The slope of the power becomes steep as $\Omega_{0}$ is lowered since
the ISW contribution transfers to the large scales. 
\\
\indent
We have performed a simple $\chi^2$ fitting analysis to the COBE DMR
band power measurements (Tegmark 1997)  (boxes) which are uncorrelated 
. We have adjusted the normalization of
the initial power to minimise the value of $\chi^2$. As shown in 
table 1, the angular power for a model with $\Omega\!=\!0.1$ 
is still within the acceptable range. The apparent primordial spectral 
index is approximately $n\!=\!1.6$ for $\Omega\!=\!0.1$.
\begin{table}
\begin{center}
\begin{tabular}{ccccccc}  
\\ \hline\hline
$\Omega_0$ & 0.1  & 0.2 & 0.3 & 0.4 & 0.5 & 0.6 \\ \hline
$\chi^2$   & 10.6 & 6.29 & 6.42 & 4.21 & 4.33 & 5.20 \\ 
$Q$   &0.15 & 0.51 & 0.49 & 0.76 & 0.74 & 0.64 \\ \hline
\end{tabular}
\caption{$\chi^2$ and the probability $Q$ that 
$\chi^2$ should exceed a particular value by chance assuming that
$\chi^2$ obeys the chi-square distribution with 7 degrees of freedom.
}
\label{tab:chisquare}
\end{center} 
\end{table}
\section{CONCLUSIONS}
 Thus the Thurston models with 
$\Omega\!\ge\!0.1$ are not constrained by the angular power spectrum
from the COBE data, which confirms the preliminary result by 
one of the author (Inoue 1999b).    
The peak at $l\!\sim\!4$ in the COBE data may be merely 
the coincidence due to the large cosmic variance but it is interesting 
that a model with $\Omega_{0}\!\sim\!0.6$ has the first peak in this scale.
Consequently, the Thurston models agree well with the COBE data than 
any FRW models.
The similar conclusion that the constraints $\Omega\!\ge\!0.3$
for an orbifold model with volume $0.7173068 R_{curv}^3$ have been
obtained in (Aurich 1999).
Although orbifolds have singular points, the behavior 
of eigenmodes for orbifolds 
is expected to be similar to that 
of manifolds. Therefore, the 
result for an orbifold model supports our conclusion. 
\\
\section*{Acknowledgments}
We would like to thank Dr. Jeff Weeks and the Geometry
Center in University of Minnesota for providing us the data of CH
spaces and Dr. Neil J. Cornish for useful comments.
The numerical computation in this work was carried out by VPP 800 
at the Data Processing Center in Kyoto University.  
K.T. Inoue is supported by JSPS Research Fellowships 
for Young Scientists, and this work is supported partially by 
Grant-in-Aid for Scientific Research Fund (No.9809834, No.11640235).

\end{document}